\documentclass[aps,prb,twocolumn,floatfix,amssymb,superscriptaddress]{revtex4}

\usepackage{amsmath}
\usepackage{amsxtra}
\usepackage{graphicx}
\usepackage{epstopdf}

\begin{document}

\title{Magnetic field induced mixed-level Kondo effect in two-level systems}   
\author{Arturo Wong}
\affiliation{Department of Physics and Astronomy and Nanoscale and Quantum Phenomena Institute, Ohio University, Athens, Ohio 45701, USA}
\affiliation{Departamento de F\'isica Te\'orica, Centro de
Nanociencias y Nanotecnolog\'ia, Universidad Nacional Aut\'onoma
de M\'exico, Apdo.\ Postal 2681,
22800 Ensenada, Baja California, M\'exico}
\author{Anh T. Ngo}
\affiliation{Materials Science Division, Argonne National Laboratory, Argonne, Illinois 60439, USA}
\author{Sergio E.\ Ulloa}
\affiliation{Department of Physics and Astronomy and Nanoscale and Quantum Phenomena Institute, Ohio University, Athens, Ohio 45701, USA}
                
\date{\today}          

\begin{abstract}
We consider a two-orbital impurity system with intra and inter-level Coulomb repulsion that is coupled to a single conduction channel. 
This situation can generically occur in multilevel quantum dots or in systems of coupled quantum dots.  
For finite energy-spacing between spin-degenerate orbitals, an in-plane magnetic field drives the system from a local singlet ground state to a ``mixed-level" Kondo regime, 
where the Zeeman-split levels are degenerate for opposite spin states.  We use the numerical renormalization group approach to fully characterize 
this mixed level Kondo state and discuss its properties in terms of the applied Zeeman field, temperature and system parameters.
Under suitable conditions, the total spectral function is shown to develop a Fermi level resonance, so that the linear conductance of the system 
peaks at a finite Zeeman field while it decreases as function of temperature.  These features, as well as the local moment and entropy contribution of the impurity
system are commensurate with Kondo physics, which can be studied in suitably tuned quantum dot systems.
\end{abstract} 

\maketitle

\section{Introduction}
The Kondo effect is a paradigmatic many-body phenomenon where an impurity magnetic moment is screened by a sea of conduction electrons, leading to a collective singlet (or reduced moment) ground state.\cite{Hewson:97}  This effect and its impact on the low temperature resistivity of magnetically doped metals has been studied a great deal, with significant developments in recent years. \cite{FeinAu-refs} 
Advances in nanolithography have also enabled extensive studies of this effect in the carefully controlled environments provided by quantum dots.\cite{Kouwenhoven:2001} In these systems, the Kondo effect manifests itself in high conductance values, where vanishing electrical current would be 
expected due to the classical Coulomb blockade.\cite{Goldhaber-Gordon:1998} The finite conductance is the result of electronic transport allowed through a Fermi-level resonance that appears in the density 
of states of the dot at temperatures below a characteristic scale known as the Kondo temperature. 

The supression of Kondo correlations by the Zeeman interaction that breaks the spin degeneracy of levels has been well established.\cite{Costi:2000} The spin polarization of the impurity due to the external magnetic field results in a strong reduction of the Kondo resonance near the Fermi energy. In quantum dot devices, this is signaled by a monotonic drop of the magnetoconductance for Zeeman energies comparable to the Kondo temperature.

In some quantum dot systems, however, the presence of a perpendicular magnetic field may result in non-monotonic magnetoconductance behavior.  
In particular, singlet-triplet transitions induced in quantum dots with even occupation give rise to complex structure in the linear magnetoconductance response, including maximum conductance at a finite magnetic field.\cite{Pustilnik:2003, Hofstetter:2004,Logan:09}  
Another interesting possibility is the diamagnetic modification of the energy level-spacing in a singly-occupied quantum dot with multiply nearly degenerate orbitals and negligible Zeeman splitting.  
In this case, the interplay between level shifts may also produce a maximum in the magnetoconductance at finite temperatures.\cite{Sakano:2006}

An interesting proposal by Pustilnik $et$ $al.$ \cite{Pustilnik:2000} showed that the Kondo effect can be induced by means of an in-plane magnetic field in quantum dots with even number of electrons. When the Zeeman energy is close to the single-particle level-spacing, the near degeneracy of the singlet and triplet configurations results in 
an effective anisotropic exchange interaction dominating the scattering of the itinerant electrons.  This creates a Kondo resonance at magnetic fields higher than the Kondo temperature of individual impurities, as experimentally observed in carbon nanotube dots with even occupation.\cite{Nygard:2000}

	The scaling and perturbative analysis used in Ref.\ [\onlinecite{Pustilnik:2000}] showed that the linear magnetoconductance of the system would be expected to indeed exhibit a maximum at a finite magnetic field.  However, the results presented lack quantitative and detailed information on the different regimes of such system.  
For this reason, we believe that the rich behavior anticipated in the model deserves close examination with an approach such as 
Wilson's numerical renormalization group (NRG), which has proven to provide an essentially exact description of  the subtle Kondo physics. 

It is the purpose of this work to carry out such systematic analysis of the problem focusing on the case of capacitively coupled quantum dots, which can be tuned to be in 
a regime described by this model.  Such systems have been studied recently in different experimental configurations. \cite{Weis,DGG-Amasha,David} 
We consider a two-level Anderson model subjected to an in-plane magnetic field and effectively connected to a single conduction channel in the current leads. The model is  first analyzed by means of a Feschbach projection and compared to the results obtained in Ref.\ [\onlinecite{Pustilnik:2000}].  This confirms that an effective exchange interaction couples the conduction channel with Zeeman-split levels that are degenerate for opposite spin states.  We then proceed with numerical renormalization group calculations of the original model. Thermodynamical quantities demonstrate that the ground state of the system is indeed a Kondo singlet when the Zeeman energy is close to the level-spacing, as it transitions away from a local singlet with no correlations with the lead electrons.  The calculated components of the spectral function show that the Kondo effect builds up with a mixture of different levels with opposite spin states. As a result, the magnetoconductance exhibits 
drastic non-monotonic behavior.  As we show, it is possible to estimate the level-spacing energy as well as the inter-level Coulomb repulsion through transport measurements.  The characteristic Kondo temperature of the mixed-level system is in general larger than for the respective single-impurity Anderson model, but also tunable by adjusting the level-spacing.  

\section{Model and approach}

      Our starting point is the two-level Anderson model\cite{Zitko:06,Logan:09} in the presence of an in-plane magnetic field,\cite{Wright:10}
\begin{equation}\label{fullH}
H=H_C+H_{T}+H_L \, .
\end{equation}
\noindent Here $H_C=\sum_{{\bf k},\sigma}\epsilon_{{\bf k}}c^{\dagger}_{{\bf k},\sigma}c_{{\bf k},\sigma}$ describes the (single) conduction channel, with $c^{\dagger}_{{\bf k},\sigma}$ as the creation operator of a spin-$\sigma$ electron with energy $\epsilon_{\bf k}$.  The second term in \eqref{fullH} describes 
the tunneling of electrons between the conduction band and the Anderson levels,   

\begin{equation}\label{hopH}
H_{T}= \sum_{i,{\bf k},\sigma}V_{i}(d^{\dagger}_{i,\sigma}c_{{\bf k} \sigma}+c^{\dagger}_{{\bf k},\sigma}d_{i,\sigma}) \, ,
\end{equation}

\noindent where  $d^{\dagger}_{i,\sigma}$ is the creation operator of a spin-$\sigma$ electron at level $i$ (=1,2).   For simplicity we consider real ${\bf k}$-independent couplings $V_{i}$. The impurity-conduction band tunneling is measured via the hybridization strength $\Gamma_i=\pi\rho V_i^2$, where a constant density of states $\rho=1/(2D)$ with half bandwidth $D$ is assumed for the leads.    

  Finally, $H_L$ describes the Anderson levels {\it per se} and reads
\begin{equation}\label{dotsH}
 H_L=\sum_{i}(\varepsilon_{i}n_i+U_i n_{i,\uparrow} n_{i,\downarrow})+U^{\prime}n_1n_2+ B S_z \, ,
\end{equation}
\noindent with $\varepsilon_{i}$, $U_i$ and $n_i=\sum_{\sigma}n_{i,\sigma}=\sum_{\sigma} d^{\dagger}_{i,\sigma}d_{i,\sigma}$ being respectively the orbital energy, the Coulomb interaction and the electron number of level $i$. The Coulomb repulsion between different levels is denoted by $U^{\prime}$.  The last term in \eqref{dotsH} takes into account the Zeeman energy with magnitude $B$, proportional to the external magnetic field, and $S_z=\frac{\ 1}{\ 2}\sum_{i,\sigma,\sigma^{\prime}} d^{\dagger}_{i,\sigma}\sigma^z_{\sigma,\sigma^{\prime}}d_{i,\sigma^{\prime}}$ is the total $z$-projection of the spin operator of the quantum dots (impurities).  We have assumed identical $g$-factors for both levels, with $g^*=\mu_B=1$, and neglect the Zeeman coupling of electrons in the conduction band.  

	In order to parameterize the energy levels it is convenient to introduce the level-spacing $\Delta=\varepsilon_1-\varepsilon_2>0$, as well as the average on-site energy $\varepsilon_{ave}=(\varepsilon_1+\varepsilon_2)/2$. The energy levels are then given by $\varepsilon_1=\varepsilon_{ave}+\Delta/2$ and $\varepsilon_2=\varepsilon_{ave}-\Delta/2$.  The described two-level Anderson model can be realized in a parallel double quantum dot device,\cite{Weis,DGG-Amasha,David} where delicate tunability of $\Delta$ and other parameters can be achieved.  

Different occupancy regimes in such system have also been considered before.  When the Zeeman coupling at the Anderson levels is neglected, the Hamiltonian \eqref{fullH} exhibits a quantum phase transition separating strong-coupling and underscreened-Kondo phases.\cite{Logan:09} The implications of a finite Zeeman splitting on this quantum phase transition, as well as in the universal behavior of the system, have been addressed in Ref.\ [\onlinecite{Wright:10}].  
As concluded in that work, the presence of magnetic fields destroys the quantum phase transition, replacing it by a smooth crossover.  
That study considered transitions going from occupancy ($\langle n_1 \rangle=1$, $\langle n_2 \rangle=1$) to ($1$, $0$).  
Note that our focus is on a different regime of double occupancy, as explained below.

\begin{figure}
\centerline{\includegraphics[width=2.5 in]{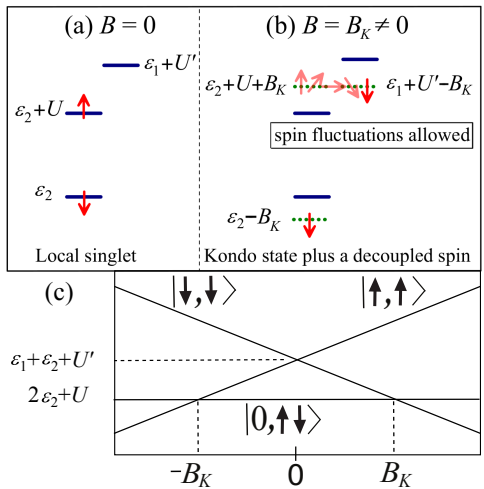}}
\caption{(Color online). Schematic representation of the level configuration considered in the present work. a) In absence of magnetic field, a local singlet develops in level 2.  A necesary condition for this situation is $\varepsilon_1+U^{\prime}>\varepsilon_2+U$. b) For a non-zero magnetic field, the levels spin degeneracy is destroyed.  However, an effective spin degeneracy can be reconstructed at a certain value of the magnetic field, by mixing the spin up and down states of different levels, allowing Kondo correlations of the impurity system with the surrounding reservoir at finite field.  The resulting spin fluctuations are indicated by the up/twisting/down red arrows. 
c) Alternative description of the scenario under study.  At $B=B_K$ (or $-B_K$), a degeneracy between the local singlet and the $S_z=-1$ (or $S_z=+1$) triplet state (which decreases its energy with $B$), gives rise to Kondo correlations.}
\label{Fig1}
\end{figure}

	Throughout this paper, the two-level system is assumed to have double occupancy and to satisfy the condition $\Delta > U-U'$ or $\varepsilon_1+U^{\prime}>\varepsilon_2+U$ (Fig.\ \ref{Fig1}), rather generically satisfied in experimental systems.  
	At zero magnetic field  [Fig.\ \ref{Fig1}(a)], this condition results in both particles being locked in a local singlet state, screening each other and preventing Kondo correlations with the leads, regardless of the levels' spin degeneracy (notice Fig.\ \ref{Fig1} shows only one possible configuration). 
The state with single occupancy of each level is suppressed at zero field by an energy difference with the local singlet given by $(2\varepsilon_{ave} + U') - (2\varepsilon_2 + U) = \Delta - (U-U') >0$.
	
	As previously discussed, a high enough parallel magnetic field suppresses the spin fluctuations in the Zeeman-split levels, which is known to destroy Kondo correlation in the single-impurity Anderson model (SIAM). \cite{Costi:2000} However, in the two-level model, increasing the magnetic field up to a value given by
\begin{equation}\label{max}
 B_K \equiv \frac{\Delta+U^{\prime}-U}{\ 2} \, ,
\end{equation}
\noindent gives rise to a spin-degenerate energy-level, built by opposite spin states of $\varepsilon_1$ and $\varepsilon_2$ [see Fig.\ \ref{Fig1}(b)], allowing spin-fluctuations in the system.  Under this scenario, the physics of the system is governed by Kondo interactions {\em restored} by the Zeeman field: the ground state is a many-body singlet, and the spectral functions develops a temperature-dependent resonance at the Fermi level.  Further increasing the magnetic field beyond $B_K$ suppresses Kondo correlations and the ground state becomes a frozen spin-integer object. 

	 This {\it sui generis} Kondo effect is alternatively described in Fig. 1 (c).  As shown by this panel, at zero magnetic field the ground state is a local singlet (both particles located at level 2), while the triplets are the lowest excited states. When the magnetic field is increased, the triplet state $S_z=-1$ decreases in energy, becoming degenerate with the local singlet state at $B=B_K$ (or degenerate with $S_z=+1$ at $B=-B_K$).  Since these states differ form each other by a spin flip and a charge transfer from one orbital to the other, the degeneracy allows the formation of the Kondo effect discussed in this paper.  
\footnote{We notice that the interplay between orbital and spin degeneracies can give rise to more unusual Kondo phenomena, as a consequence of SU(4) symmetry setting in, as reported in carbon nanotubes, \cite{Jarrilo:2005,Choi:2005} and double quantum dots.\cite{Borda:2003, DGG-Amasha}  Such degeneracy is not present in the regime we study here.}

	In the following section we present a Schrieffer-Wolff transformation of the Hamiltonian \eqref{fullH} to show that the scattering mechanism for itinerant electrons is dominated by exchange interactions in presence of a magnetic field.  We compare our results to those obtained in Ref.[\onlinecite{Pustilnik:2000}] and we use them to describe the NRG calculations presented in further sections.
	
	\section{Low energy effective model}

	By using a Feshbach projection-operator formalism on Eq.\ \eqref{fullH}-\eqref{dotsH}, we integrate out the states with $\langle n_1+n_2\rangle\ne2$.  The resulting effective Hamiltonian reads 
\begin{equation}\label{effectiveH}
H^{eff}=H_C+\sum_{i\sigma i^{\prime}\sigma^{\prime}}(H^{ex}_{i\sigma i^{\prime}\sigma^{\prime}}+H^{ps}_{i\sigma i^{\prime}\sigma^{\prime}})(1-\delta_{ii^{\prime}}\delta_{\sigma\sigma^{\prime}})\bar{n}_{i\sigma i^{\prime}\sigma^{\prime}}.
\end{equation}

\noindent The factor $\bar{n}_{i\sigma i^{\prime}\sigma^{\prime}}= n_{\bar{i}\bar{\sigma}}\delta_{ii^{\prime}}+ n_{i\bar{\sigma}}(1-\delta_{ii^{\prime}})$, with $\bar{i}\ne i$ and $\bar{\sigma} \ne \sigma$, guarantees the double occupancy of the levels (see below).  The first term in the sum of  \eqref{effectiveH} is an anisotropic exchange interaction
\begin{equation}
\begin{split}\label{exchange}
H^{ex}_{i\sigma i^{\prime} \sigma^{\prime}}&={\cal J}^{\perp}_{i\sigma i^{\prime}\sigma^{\prime}}(s_{\sigma\sigma^{\prime}}^{x}{\cal S}_{i\sigma i^{\prime}\sigma^{\prime}}^{x}+s_{\sigma\sigma^{\prime}}^{y}{\cal S}_{i\sigma i^{\prime}\sigma^{\prime}}^{y})\\
&+{\cal J}^z_{i\sigma i^{\prime}\sigma^{\prime}} s_{\sigma\sigma^{\prime}}^{z}{\cal S}_{i\sigma i^{\prime}\sigma^{\prime}}^{z}+{\cal J}^c_{i\sigma i^{\prime}\sigma^{\prime}} \rho_{\sigma\sigma^{\prime}}{\cal S}_{i\sigma i^{\prime}\sigma^{\prime}}^{z} \, ,
\end{split}
\end{equation}

\noindent where the various coefficients are given by ${\cal J}^{\perp}_{i\sigma i^{\prime}\sigma^{\prime}}=2V_iV_{i^{\prime}}J_{i\sigma i^{\prime}\sigma^{\prime}}$, ${\cal J}^{z}_{i\sigma i^{\prime}\sigma^{\prime}}=(V_i^2+V_{i^{\prime}}^2)J_{i\sigma i^{\prime}\sigma^{\prime}}/2$, ${\cal J}^{c}_{i\sigma i^{\prime}\sigma^{\prime}}=(V_i^2-V_{i^{\prime}}^2)J_{i\sigma i^{\prime}\sigma^{\prime}}/2$, and

\begin{equation}
J_{i\sigma i^{\prime}\sigma^{\prime}}=\frac{\ 1}{\ \varepsilon_i+U_i+U^{\prime}+\sigma B}-\frac{\ 1}{\ \varepsilon_{i^{\prime}}+U^{\prime}+\sigma^{\prime} B} \, .
\end{equation}

\noindent We have also defined the spin operators ${\cal S}^{z}_{i\sigma i^{\prime}\sigma^{\prime}}=\frac{\ 1}{\ 2}(d^{\dagger}_{i\sigma}d_{i\sigma}- d^{\dagger}_{i^{\prime}\sigma^{\prime}}d_{i^{\prime}\sigma^{\prime}})$ and ${\cal S}^{+}_{i\sigma i^{\prime}\sigma^{\prime}}={\cal S}^{x}_{i\sigma i^{\prime}\sigma^{\prime}}+ i{\cal S}^{y}_{i\sigma i^{\prime}\sigma^{\prime}}=d^{\dagger}_{i\sigma}d_{i^{\prime}\sigma^{\prime}}$. Similarily, $s^{z}_{\sigma\sigma^{\prime}}=\frac{\ 1}{\ 2}\sum_{\bf{k}\bf{k}^{\prime}}
(c_{\bf{k}\sigma}^{\dagger}c_{\bf{k}^{\prime}\sigma}-c_{\bf{k}\sigma^{\prime}}^{\dagger}c_{\bf{k}^{\prime}\sigma^{\prime}})$, $s^{+}_{\sigma\sigma^{\prime}}=s^{x}_{\sigma\sigma^{\prime}}+is^{y}_{\sigma\sigma^{\prime}}=\sum_{\bf{k}\bf{k}^{\prime}} c^{\dagger}_{\bf{k}\sigma}c_{\bf{k}^{\prime}\sigma^{\prime}}$ and $\rho_{\sigma\sigma^{\prime}}=\frac{\ 1}{\ 2}\sum_{\bf{k}\bf{k}^{\prime}}(c_{\bf{k}\sigma}^{\dagger} c_{\bf{k}^{\prime}\sigma}+c_{\bf{k}\sigma^{\prime}}^{\dagger} c_{\bf{k}^{\prime}\sigma^{\prime}})$.  Note that the last term in \eqref{exchange} couples the conduction-band charge degrees of freedom to the levels' spin $z$-component, and it vanishes for $V_i^2=V_{i^{\prime}}^2$.  In addition to the exchange interaction, Eq.\ \eqref{effectiveH} includes a potential scattering term
\begin{equation}
H^{ps}_{i\sigma i^{\prime} \sigma^{\prime}}=\frac{\ {\cal K}^z_{i\sigma i^{\prime}\sigma^{\prime}}}{\ 4}s_{\sigma\sigma^{\prime}}^{z}+\frac{\ {\cal K}^{c}_{i\sigma i^{\prime}\sigma^{\prime}}}{\ 4}\rho_{\sigma\sigma^{\prime}}
\end{equation}

\noindent with ${\cal K}^{z}_{i\sigma i^{\prime}\sigma^{\prime}}=(V_i^2+V_{i^{\prime}}^2)K_{i\sigma i^{\prime}\sigma^{\prime}}$, ${\cal K}^{c}_{i\sigma i^{\prime}\sigma^{\prime}}=(V_i^2-V_{i^{\prime}}^2)K_{i\sigma i^{\prime}\sigma^{\prime}}$, and

\begin{equation}
K_{i\sigma i^{\prime}\sigma^{\prime}}=-\frac{\ 1}{\ \varepsilon_i+U_i+U^{\prime}+\sigma B}-\frac{\ 1}{\ \varepsilon_{i^{\prime}}+U^{\prime}+\sigma^{\prime} B}.
\end{equation}

    Equation \eqref{effectiveH} is the main result of this section.  It contains the effective Hamiltonian obtained in Ref.\ [\onlinecite{Pustilnik:2000}] by a Schrieffer-Wolff transformation, plus additional terms describing different types of  Kondo correlations, as explained below.   Notice we have chosen to write the various  coefficients in a slightly different manner.  Here, $K_{i\sigma i^{\prime}\sigma^{\prime}}$ vanishes at the particle-hole symmetric point (as it occurs in the SIAM) in contrast to the corresponding term in Ref.\ [\onlinecite{Pustilnik:2000}], which does not. Notice also that our ${\cal J}^{z}$ parameter is one half of the respective term in the same reference.
    
    The Hamiltonian \eqref{effectiveH} contains three different types of exchange interactions. The contribution from each one strongly depends on the magnetic field and on the levels' configuration.  The simplest components are given by the $i=i^{\prime}$ terms, which are the usual Kondo coupling between the levels and the conduction band.  This well-known Kondo interaction dominates $B=0$, in which case the $i\ne i^{\prime}$ terms are negligible.
    
    In contrast, the terms with $i\ne i^{\prime}$, $\sigma=\sigma^{\prime}$ represent a {\it charge Kondo effect} where the Kondo correlations come from the orbital degeneracy for a fixed spin state, rather than from spin fluctuations within a single orbital as in the former case.  This kind of exchange interaction is restricted to $\varepsilon_1=\varepsilon_2$ and persists in a finite magnetic field.  However, the  configuration when the ground state is dominated by the triplet state $S_z=-1$ (or $+1$) prevents the described charge correlations.
    
    The last type of exchange interaction is the main focus of this paper.  It is given by the $i\ne i^{\prime}$, $\sigma\ne\sigma^{\prime}$ terms.  As described in the previous section, this requires the different levels to be degenerate for opposite spin states, which can only be achieved by a finite magnetic field. For the particular case depicted in Fig.\ 1, ($i,\sigma=2,\uparrow$, $i^{\prime},\sigma^{\prime}=1,\downarrow$) the exchange interaction at $B=B_K$ gives
    
 \begin{equation}\label{JforFig2}
J_{2\uparrow 1\downarrow}=\frac{\ 2}{\ 2\varepsilon_{ave}+U_2+3U^{\prime}}-\frac{\ 2}{\ 2\varepsilon_{ave}+U_2+U^{\prime}}
\end{equation}
    
\noindent and $K_{2\uparrow 1\downarrow}$ can be obtained by reversing the sign of the first term in the last equation.  Note also that $\bar{n}_{2\uparrow1\downarrow}=n_{2\downarrow}$, $i.e.$ the process requires a spin-down particle to be located at $\varepsilon_2$ [see fig. 1 (b)].

	Equation \eqref{JforFig2} provides an analytic expression to estimate the exchange couplings of the mixed-level Kondo effect in terms of the model parameters. Note that for $V_1=V_2=V$, and at the particle-hole symmetry point $\varepsilon_{ave} = -U/2-U^{\prime}$, the exchange couplings reduce to the simple expressions ${\cal J}^z_{2\uparrow 1\downarrow}=4V^2/U^{\prime}={\cal J}^{\perp}_{2\uparrow 1\downarrow}/2$, close to the SIAM's expression of the exchange coupling.  Note that both ${\cal J}^z$ and ${\cal J}^{\perp}$ are independent of the levels' splitting $\Delta$.  From this, one may infer that the mixed-level Kondo temperature will remain the same for different values of the detuning.  However this is not the case, as we further show in this paper using NRG calculations.

\section{Numerical renormalization approach}

       In order to provide a quantitative analysis of the behavior described above, we have used the numerical renormalization group (NRG) approach.\cite{BullaRMP} The NRG method has proven to be a powerful tool for studying quantum impurity systems in a non-perturbative scheme, as the technique allows the calculation of system properties with high accuracy.  In particular, the calculation of the impurities' contribution to the entropy $S_{\text{imp}}$ as a function of temperature $T$, permits the identification of effective degrees of freedom of the system. In the usual fashion it can be computed as,
\begin{equation}
 S_{\text{imp}}(T)= \frac{\ (E-F)}{\ T}- \frac{\ (E-F)_0}{\ T} \, ,
\end{equation}

\noindent where $E=\langle H \rangle$ is the mean energy (relative to the ground state) of the entire system, $F=-k_B T\ln Z$ is the free energy and $Z=\sum e^{-H/T}$ is the partition function. The symbol $\langle...\rangle$ denotes thermal average and the subscript $0$ refers to the situation when no impurities are present. 

	Further insight of the interaction between the Anderson levels and the external magnetic field can be gained by studying the contribution of the quantum dot system to the magnetization and susceptibility, $M_{\text{imp}}$ and $T\chi_{\text{imp}}$, respectively.  These quantities can be calculated as
\begin{equation}
M_{\text{imp}}(T)=\langle S_z \rangle - \langle S_z \rangle_0 \, ,
\end{equation}

\noindent and
\begin{equation}
T\chi_{\text{imp}}(T)=(\langle S_z^2 \rangle - \langle S_z \rangle^2)-(\langle S_z^2 \rangle - \langle S_z \rangle^2)_0 \, ,
\end{equation}

\noindent where $S_z$ is the $z$ component of the total system spin.

	By means of the NRG procedure it is also possible to calculate transport properties. At low bias, electron transmission described by a generalized Landauer formula \cite{Meir:92} gives a linear conductance

\begin{equation}
g=\frac{2e^2}{h}\int d\omega \biggr(\frac{-\partial f}{\partial\omega}
\biggr) \, \pi \sum_{i,j}\sqrt{\Gamma_i \Gamma_j} \, A_{ij}(\omega,T),
\end{equation}

\noindent where $f(\omega,T)$ is the Fermi function, $A_{ij}(\omega)=-\pi^{-1} \, \text{Im} \, G_{ij}(\omega)$ are the componentes of the total spectral function $A= \sum_{i,j} A_{ij}$, and $G_{ij}(t)=-i\theta(t)\langle\{d_{i,\sigma}^{\dagger}(t), \, d_{j,\sigma}^{\dagger}(0) \}\rangle$
is the retarded Green's function.

The following sections are devoted to prove that at $B=B_K$, the ground state of the two-level Anderson system is indeed a many-body (Kondo) singlet state and
to explore the behavior of the system for different $B$ fields and level spacing $\Delta$, 
by analyzing the thermodynamical and transport properties described above.  Calculations were performed considering $H/D=6.01\times 10^{-5}$ as a unit of energy, which corresponds to a half of the Zeeman splitting for a magnetic field of 1 T in a  $GaAs$ system\cite{Tai-Min:09} with $g^*\mu_B=12.02$ $\mu$eV/T  and $D=100$ meV. For simplicity we set $U=U^{\prime}=30H$, $\Gamma_1=\Gamma_2=\Gamma \ll U$ such that the system is well inside the Kondo regime, and restrict the calculations to the particle-hole symmetric point.  For this regime, as we discussed in the previous section, the potential scattering terms vanish, together with the coupling between the impurity spin and the conduction-channel charge degrees of freedom. $\Delta$ and $\Gamma$ values used are reported in each case below.

\section{Thermodynamics}

\begin{figure}
\centerline{\includegraphics[width=3.4 in]{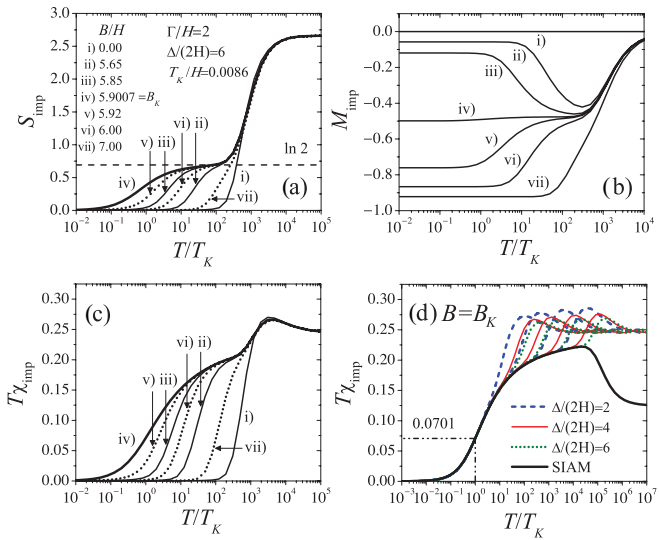}}
\caption{Temperature dependence of the impurities' contribution to (a) the entropy, (b) magnetization and (c) susceptibility for different strengths of the magnetic field. At fixed $B=0$, curves i),  reducing temperature makes the entropy go directly from $\text{ln}(4^2)$ to $\text{ln}(1)$, the susceptibility from $0.25$ to zero while the magnetization remains null, as expected for a local singlet-state. As the magnetic field turns on, curves ii) and iii), it allows the involvement of different levels with other spin values, as the entropy and susceptibility exhibit plateaus near $S_{\text{imp}}=\text{ln}(2)$ and $T\chi_{\text{imp}}=0.20$, respectively, before dropping to zero at low temperature, signaling the quenching of a doublet ground state.  This leaves behind a frozen spin-1/2 object with $M_{\text{imp}}\approx-1/2$ at $B=B_K$, curves iv).  Further increasing the magnetic field results in a frozen spin-integer object with $M_{\text{imp}}\approx-1$, curves v)-vii). Panel (d) shows the  susceptibility for different values of $\Gamma$, $\Delta$ and fixed $B=B_K$, demonstrating the universal behavior around $T_K$ and below; for each $\Delta$, $\Gamma/H=1.25, 1.6, 2.2$ and $3$.}
\label{Fig2}
\end{figure}

	We start our numerical analysis by discussing the thermodynamical behavior of the two-level system in the presence of Zeeman interaction. In Fig.\ \ref{Fig2} we show the temperature dependence of the Anderson impurities' contribution to a) the entropy, b) the magnetization and c) the susceptibility, considering $\Delta/(2H)=6$, $\Gamma=2H \ll U$, and several values of $B$.  In all three panels, the horizontal axis has been rescaled by the Kondo temperature $T_K=0.0086H$, which was extracted from the spin susceptibility calculations (see below).  At high temperature, the system is in the free-orbital regime, and its ground state is $4^2$-degenerate, $i.e.$ $S_{\text{imp}}=\text{ln} (16)$, with zero magnetization and $T\chi_{\text{imp}}=1/4$ (a contribution of $1/8$ from each impurity level).  In the absence of an external magnetic field [$B=0$, curves labelled i)], reducing the temperature results in a suddently flat (zero) magnetization and accompanying monotonic decrease of the entropy towards $S_{\text{imp}}=\text{ln} (1)$, in accordance with the existence of a local singlet ground state.  The susceptibility displays a similar behavior to that of the entropy, with a slight bump at $T/T_K\gtrsim10^3$, characteristic of two level systems.\cite{Logan:09}

	At finite magnetic fields, $B \ll B_K$ [curves ii) and iii)], the magnetization exhibits a dip with minimum near $M_{\text{imp}}=-1/2$.  This drop results from the gap closing between the $\varepsilon_2$ spin-up and the $\varepsilon_1$ spin-down states by the combined effect of temperature and Zeeman splitting, creating a mixed level where spin fluctuations are allowed in the system.  This dynamics is reflected in the respective entropy curves as a plateau near $S_{\text{imp}}=\text{ln} (2)$, as well as in the susceptibility plot with a plateau around $T\chi\approx0.20$, signaling the existence of a doublet state (a local-moment fixed point), although here it is associated with a four-state manifold of the doubly-occupied two-level system.  Since the system is doubly occupied, this leaves behind a frozen spin-1/2 object, resulting in $M_{\text{imp}}\approx-1/2$.  However in this situation, the gap between the opposite-spin levels can not be bridged solely by the Zeeman splitting, and at very low $T$ the two particles lock in a local-singlet state, which is barely affected by the magnetic field, resulting in $M_{\text{imp}}\approx 0$ and zero entropy and susceptibility at low temperatures. 

	As expected, further increasing magnetic field yields interesting results.  The resonance between opposite spin levels can be achieved by a high enough magnetic field, as shown by curve iv) in all three panels for $B=B_K$.  In this case, one particle is spin-fluctuating while the other is magnetically frozen, so that the magnetization remains at $M_{\text{imp}}=-1/2$ down to low temperature.  On the other hand, the entropy shows a smooth decrease from $S_{\text{imp}}=\text{ln} (2)$ to $S_{\text{imp}}=\text{ln} (1)$ and the susceptibility from $T\chi\approx0.2$ to $T\chi=0$, both at the Kondo temperature.  This transition from doublet (unscreened magnetic impurity or
local-moment fixed point) to a nonlocal singlet state  
(strong-coupling fixed point) is a clear signature of Kondo state formation, and confirms one of the main points of this paper. According to relation \eqref{max}, the Kondo effect is expected to occur at $B/H=6$, which is in agreement with the value of $B_K=5.9007 \gg T_K$ found from susceptibility calculations (see below), except for the small anticipated shift produced by the renormalization of the levels due to the coupling to the leads and Coulomb interaction.
	
	If the intensity of the magnetic field is slightly above $B_K$ [v) and vi) curves], the magnetization exhibits a plateau near $M_{\text{imp}}=-1/2$ and $S_{\text{imp}} \simeq \text{ln}(2)$ for $T \simeq 10T_K$, which can be seen again in terms of the quasi-degeneracy of the opposite spin states of the levels, due to the combination of thermal and magnetic effects, as described above.  However, now the spin-down states of both levels are energetically more favorable and the magnetization reaches values below $-3/4$ at very low $T$.  At very high magnetic fields [curve vii)] the entropy and magnetization of the system flow directly from $S_{\text{imp}}=\text{ln} (16)$ to $S_{\text{imp}}=\text{ln} (1)$ and from $M_{\text{imp}}=0$ to $M_{\text{imp}}=-1$ exhibiting a ground state with $S_z=1$. The susceptibility goes directly from $T\chi_{\text{imp}}\approx0.20$ to $T\chi_{\text{imp}}=0$ at $T=B$, supporting the same conclusion.

	The behavior of the susceptibility for different values of $B$ as exhibited by Fig.\ 2(c), allows us to determine $T_K$ and $B_K$ of the system in the following way.  First, one defines a magnetic field dependent temperature $T_0(B)$ such that $T_0\chi(T_0)=0.0701$, used as the standard condition for the Kondo temperature.\cite{Krishna-Murthy:1980}  As shown in panel c), we expect that the minimum of the function $T_0(B)$ will be precisely at the point $T_K=T_0(B=B_K)$.  In order to validate this criterion, Fig.\ 2(d) explores the universality of this mixed-level Kondo effect. Here we plot the contribution of the system impurities to the susceptibility as a function of $T/T_K$, for $\Delta / (2H) = 2$ (blue-dashed), $4$ (red-solid) and $6$ (green-dotted), and different values of $\Gamma/H$ (=$1.25$, $1.6$, $2.2$ and $3$), all of them at $B=B_K$.  All these curves are compared with the susceptibility obtained for the SIAM (black-solid).  One can appreciate that the two-level behavior dominates at high temperature.  However, all the curves collapse into a single curve and demonstrate universal behavior over a wide range of temperatures around and below $T_K$, validating our definition of the Kondo temperature for the mixed-level Kondo effect.  This figure further confirms that at $B=B_K$, the ground state of the system is indeed a many-body Kondo singlet.

\begin{figure} 
\centerline{\includegraphics[width=2.5 in]{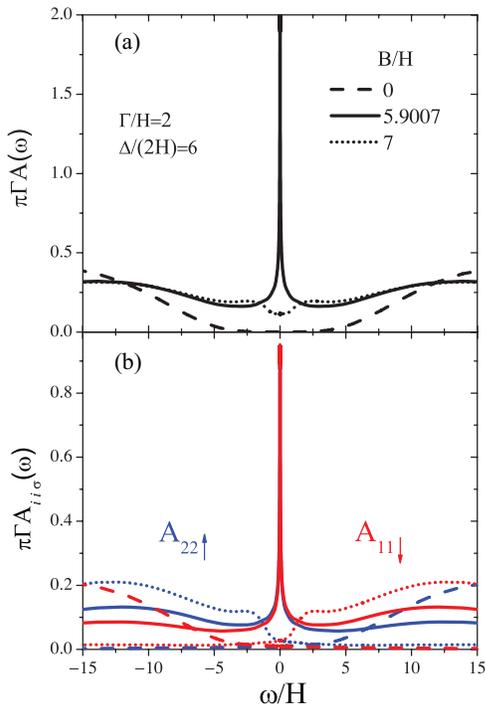}}
\caption{(Color online). (a) Total spectral function for different values of $B$. At $B=B_K=5.9007$, the spectral function exhibits a sharp peak at the Fermi level, commensurate with Kondo physics, and absent for other fields. (b) Components of the spectral function showing that at $B=B_K$ the main contributions to the Kondo peak come from the diagonal parts of $A(\omega)$ with opposite spins.}
\label{Fig3}
\end{figure}

\begin{figure} 
\centerline{\includegraphics[width=3.4 in]{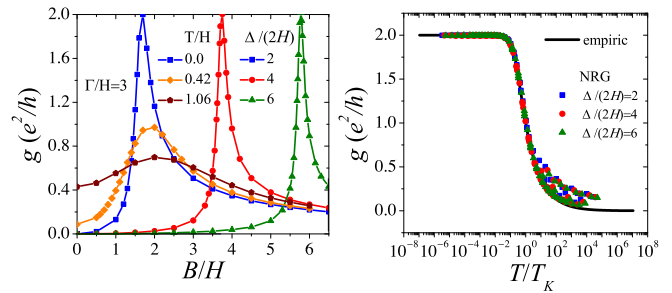}}
\caption{(Color online). Conductance $g$ as a function of $B$ (left panel) and $T$ (right panel) for different values of $\Delta$.  The magnetoconductance displays a maximum structure near $B=B_K$ ($\simeq \Delta/2$, as $U=U'$ here), as described by Eq.\ \eqref{max}. The conductance decreases by increasing $T$ above the Kondo temperature, as expected in the Kondo effect.} 
\label{Fig4}
\end{figure}

\section{Dynamics and transport}

	The emergence of a sharp peak at the Fermi level in the spectral function is a another well established signature of Kondo physics. Figure \ref{Fig3}(a) shows the normalized total spectral function $\pi\Gamma A(\omega)$ at $T=0$, for the same system parameters as in Fig.\ \ref{Fig2}(a), and three different values of $B$.  The appearance of a Kondo resonance when $B=B_K$ is evident, supporting the existence of a magnetic field induced mixed-level Kondo state.  This is in contrast to larger or lower values of $B$, where no peak at the Fermi level develops.   

	It is illustrative to analyze the components of the Anderson levels' spectral function.  Figure \ref{Fig3}(b) shows the (normalized) diagonal terms of the total spectral function with opposite spins, $\pi\Gamma A_{11,\downarrow}(\omega)$ and $\pi\Gamma A_{22,\uparrow}(\omega)$.   Note that at $B=B_K$, these spectral function components form the main contributions to the Kondo peak, as discussed above.  The rest of the components (not shown) display no Fermi-level resonance.  This further confirms that the Kondo effect is taking place due to the mixture of levels with opposite spin, creating a hybrid level with spin-degeneracy.  For magnetic fields different to $B_K$ the components of the spectral function show no more structure than the Hubbard bands at $\pm U/2$.

	The mixed-level Kondo effect can be experimentally confirmed through conductance measurements.  In Fig.\ \ref{Fig4} (left panel) we show the calculated linear conductance $g$ against magnetic field, for a few values of $\Delta$ and different temperatures. In this case we consider $\Gamma=3H$, to better highlight conductance features, with the rest of parameters as before. Let us first focus on the $T=0$ conductance.   The most important feature depicted in this panel is the peak structure clearly shown by the magnetoconductance at a {\em finite} $B$ field.  This is in sharp contrast with the monotonic drop of $g(B)$ that accompanies the Kondo effect in the SIAM, and typically seen in experiments.

	From the discussion above, the mixed-level Kondo effect would be expected to appear in the conductance only at a magnetic field $B=B_K$, as the Kondo peak in the spectral function allows high electronic transport at vanishing bias.  Lowering or increasing the magnetic field below or above $B_K$ results in a rapid drop in the magnetoconductance (made smoother at finite temperature), due to the crossover from the Kondo singlet state to either the local singlet or the magnetically frozen impurity.  As can be seen in the left panel in Fig.\ \ref{Fig4}, the maximum of the conductance appears near $B \simeq B_K \simeq \Delta/2$ (as $U=U'$ in this example), in agreement with Eq.\ \eqref{max}.  In other words, the position of the maximum in the magnetoconductance depends on the detuning $\Delta$. In fact,  if the condition $\varepsilon_1+U^{\prime}>\varepsilon_2+U$ is satisfied, it is possible to estimate the value of $\Delta$ by measuring the energies $E_1=\varepsilon_2$, $E_2=\varepsilon_2+U$ and $E_3=E_2+\Delta+2U^{\prime}$, which are the energies needed to add the first, second and third electron, respectively, to the double dot system.  It follows from Eq.\ \eqref{max} that $\Delta=2B_K+U-U^{\prime}$, where $B_K$ is the magnetic field at which the maximum in conductance occurs in $g$ vs. $B$ measurements. By using $U=E_2-E_1$, together with $E_3$ one obtains, 
\begin{equation}
\Delta=4B_K+3E_2-2E_1-E_3.
\end{equation}

\noindent In the same way, the inter-level Coulomb repulsion $U^{\prime}$ can be estimated as $U^{\prime}=2B_K+E_2-E_1-\Delta$.  This provides a useful methodology to experimentally estimate or confirm the energy level-spacing $\Delta$ through magnetoconductance measurements.

	We now focus on the temperature dependence of the linear transport.  The left panel in Fig.\ \ref{Fig4} also shows finite-$T$ calculations for $\Delta/(2H)=2$. For these parameters, the Kondo temperature, as extracted from the spin susceptibility, is estimated as $T_K=0.16H$.  Figure \ref{Fig4} shows a clear drop in $g$ (for the $\Delta/(2H)=2$ case) as the temperature is increased above the Kondo temperature, in agreement with Kondo physics.  The universal nature of the mixed-level Kondo effect is also confirmed in the right panel of Fig.\ \ref{Fig4}.  That panel shows $g$ vs $T/T_K$ at $B=B_K$ for the same parameters as in Fig.\ \ref{Fig2}(d), and a comparison to the empirical formula\cite{Goldhaber-Gordon:PRL1998}

\begin{equation}
g(T)=g_{max}[1+(2^{1/s}-1)(T/T_K)^2]^{-s}\, ,
\end{equation}

\noindent where $g_{max}=g(T=0)$ is the highest value of the conductance and $s$ is an adjustable fitting parameter.  This empirical equation is a good representation not only for experiments but for NRG calculations of the Kondo effect. The scaled NRG data are in good agreement with a theoretical curve (solid line) calculated using $s= 0.22$, a value expected for spin-1/2 Kondo physics.\cite{Goldhaber-Gordon:1998, Costi:1994} Small deviations from the empirical formula appear at temperatures $T\approx\Gamma$.

\section{Mixed-level Kondo temperature}

\begin{figure}[b]  
\centerline{\includegraphics[width=3.4 in]{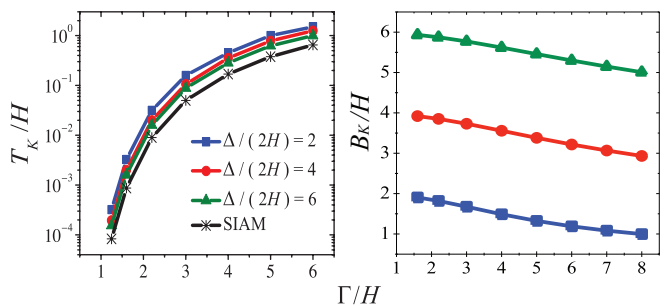}}
\caption{(Color online).  Kondo temperature (left panel), and field (right) of the mixed-level Kondo effect as function of $\Gamma$ and various $\Delta$. $T_K$ shows exponential behavior, commensurate with Kondo physics and the Haldane formula. The characteristic Kondo temperature of the mixed-level system is {\em higher} than the respective $T_K$ of the SIAM, but they get closer for larger $\Delta$ values. Right panel shows how increasing $\Gamma$ slightly decreases the required $B_K$ to produce the mixed-level Kondo ground state.}
\label{Fig5}
\end{figure}

	Since the Kondo temperature is the quantity that epitomizes a Kondo system, a natural question in the mixed-level Kondo effect is to inquire about the behavior of its $T_K$ with system parameters.  Figure \ref{Fig5} shows he Kondo temperature of the mixed-level system, as a function of $\Gamma$ for different values of $\Delta$.   As reference we also show the corresponding $T_K$ of the SIAM, for the same Coulomb interaction at particle-hole symmetry. For all the values of the level-spacing $\Delta$ considered, $T_K$ increases exponentially with $\Gamma$, in agreement with Kondo physics (and the Haldane formula). However, one expects that as $\Delta$ increases, the magnetically frozen spin-1/2 located at the lowest level interacts less with the mixed-level structure and the system resembles more the SIAM, as reflected in their respective Kondo temperatures.  Consequently, the higher the value of $\Delta$, the lower the value of $T_K$, as seen in Fig.\  \ref{Fig5}. In other words, in the mixed-level system, the level detuning can be seen as an extra parameter to control the Kondo temperature of the structure.
	
	Figure \ref{Fig5} also shows (right panel) that for larger $\Gamma$ and fixed interlevel spacing $\Delta$, the critical value of $B_K$ decreases slightly.  This behavior underscores that as the levels become effectively broader, the field required to achieve effective spin fluctuations decreases somewhat, as one would intuitively expect.

\section{Summary}

	 We have extended the study of a different type of Kondo effect,  first studied in Ref.\ [\onlinecite{Pustilnik:2000}], which arises from the mixture of different levels with opposite spins, in the presence of an in-plane magnetic field.  Examining the resulting model obtained by integrating out high energy states, we find different types of exchange possible in the system.  Moreover, our numerical renormalization results fully characterize the thermodynamics and transport properties of the system in the interesting regime of level degeneracy for opposite spin states due to the Zeeman field.
	 
	 The system under study can be seen to appear in capacitively coupled double quantum dot systems, although it may also be relevant in multilevel single-dot geometries.
	As a result of the level detuning, the magnetoconductance exhibits a sharp non-monotonic behavior, in contrast to the single impurity Anderson model. Interestingly, the doubly-occupied double quantum dot system evolves from a local singlet ground state at zero field, to a many body singlet with Kondo correlations as the Zeeman field overcomes the level spacing in the structure.
	We have shown that this phenomenon can be used to experimentally estimate the level-spacing in quantum dots.  In addition, the characteristc Kondo temperature can be modified not only by the standard parameters of the Anderson model, but also by the detuning of the quantum dot.  
	
	Finally, we should mention that the presence of sizable spin-orbit coupling in quantum dot systems may give rise to shifts in the characteristic $B_K$ fields with interesting  dependence of the Kondo conductance and temperature with applied field orientation in the plane of the dots.\cite{Sabine}

\acknowledgments
We acknowledge support from NSF grant DMR 1508325 and useful discussions with A. Kogan and his group at U. Cincinnati.

\end{document}